\documentclass[twocolumn,twoside,slac_two]{revtex4}
\usepackage{graphicx}
\usepackage{fancyhdr}
\def\lsi61{LS I +61$^{\circ}$ 303}

\begin{document}

%Title of paper
\title{Hunting for New Gamma-ray Binaries - 
Technique Development}

% Repeat the \author .. \affiliation  etc. as needed
%
% \affiliation command applies to all authors since the last
% \affiliation command. The \affiliation command should follow the
% other information

\author{Robin H.D. Corbet}
\affiliation{University of Maryland, Baltimore County, MD, USA}
\affiliation{CRESST/Mail Code 662, X-ray Astrophysics Laboratory,
NASA Goddard Space Flight Center, Greenbelt, MD 20771, USA}
\author{Matthew Kerr}
\affiliation{Department of Physics,
University of Washington, Seattle, WA 98195, USA}

\begin{abstract}
There are only a few sources that are definitely known to be gamma-ray
binaries. Two of these are listed as associations in the Fermi LAT
Bright Source List. We are developing novel techniques to extract high
signal-to-noise light curves of all cataloged Fermi sources and to
search for periodic variability using appropriately weighted power
spectra. The detection of periodic variability would be strong
evidence for the detection of a new gamma-ray binary. The LAT's
sensitivity provides the opportunity to open up completely new
discovery space for additional binary systems, potentially involving
novel astrophysics. We present here demonstrations of the sensitivity
gains obtained through the use of these techniques.

\end{abstract}

%\maketitle must follow title, authors, abstract
\maketitle

\thispagestyle{fancy}

% body of paper here - Use proper section commands
% References should be done using the \cite, \ref, and \label commands
% Put \label in argument of \section for cross-referencing
%\section{\label{}}

\section{Introduction}
At X-ray energies, the extra-solar sky is dominated by the emission
from accreting binary systems. However, at higher energies (GeV to
TeV) very few binary systems are known to be sources 
(e.g. Holder 2009). The emission
mechanisms of gamma-ray binaries are still unclear. The principal
models proposed are that either the gamma-ray emission could originate
from relativistic jets generated by accretion onto a neutron star or
black hole (``microquasars'') or be due to the interaction between the
relativistic wind coming from a pulsar and the stellar wind of its
companion.

We are investigating techniques to obtain LAT light curves with
increased signal-to-noise levels and search for periodic modulation of
the gamma-ray flux which would be a strong indicator that the
source is a binary system.
The usual way to search for periodic variability
is to use a periodogram (e.g. Scargle 1982). The sensitivity of the
periodogram for the detection of a periodic signal strongly depends on
the signal strength compared to the noise level: the probability of a
peak in the periodogram reaching or exceeding a value ``Z'' scales as
exp(-Z) when the periodogram is normalized by the variance of the data
set. Small changes in signal-to-noise can thus result in large
significance changes in a power spectrum.

\section{The Challenge}
In the optical and X-ray wavebands aperture photometry is relatively
straightforward. Although the ideal aperture to use may depend on
source brightness, the point spread function (PSF) 
typically has little energy dependence and
minor dependence on the source location in the field of view. The
signal-to-noise ratio (SNR) of a single observation will be given by:
SNR = S/(S + B)$^{1/2}$, where S is the number of source photons and B is
the number of background photons.  In the optical it is generally
relatively easy to determine the aperture at which the SNR is greatest
(e.g. Howell 1989). For the Fermi LAT the situation is more difficult
since the PSF depends strongly on energy. In addition, the background, both
from other sources and the Galactic plane, is complex.

The usual alternative to aperture photometry for LAT data is to use
maximum likelihood fitting (the equivalent of profile fitting in the
optical). However, this procedure is both very compute intensive and
can also be problematic when few or zero photons are detected in a
time bin.

\section{Approaches to the Problem}
\subsection{Optimal Aperture and Energy Range Selection}
We are developing tools to determine the ideal aperture size and
energy range to use for any source. These tools take into account the
cataloged spectrum of the source of interest, potentially
contaminating nearby sources, and emission from the Galactic plane.

Once apertures and energy ranges have been determined, aperture
photometry will then be generated for all cataloged LAT sources. We choose to
analyze all sources, not just those thought most likely to be
gamma-ray binaries. This ensures that a misclassified gamma-ray binary
will not be missed.  A large fraction of the light curves will be of
AGNs and we will provide our light curves to the community as a
general service. We use discrete Fourier Transforms and
Lomb-Scargle analyses to search for the periodic modulation that is
expected to be a signature of a gamma-ray binary.

\subsection{Photon Weighting}
Evaluating the probability that each individual photon came from a
specific source should give the maximum possible SNR. This is being
investigated for pulsars by Kerr (2009) and will later be expanded
to other types of sources.

\subsection{Multiple Apertures}
As an intermediate step between optimal apertures and photon weighting
we are also investigating the SNR gain provided by obtaining light
curves in several energy bands with different aperture sizes at each
energy band. i.e. smaller apertures at higher energies where the PSF
is smaller. These light curves can then be combined together with
appropriate weighting into an overall light curve. This technique has
some similarities to Naylor's (1989) ``optimal photometry'' method.

\begin{figure}
\includegraphics[width=65mm,angle=270]{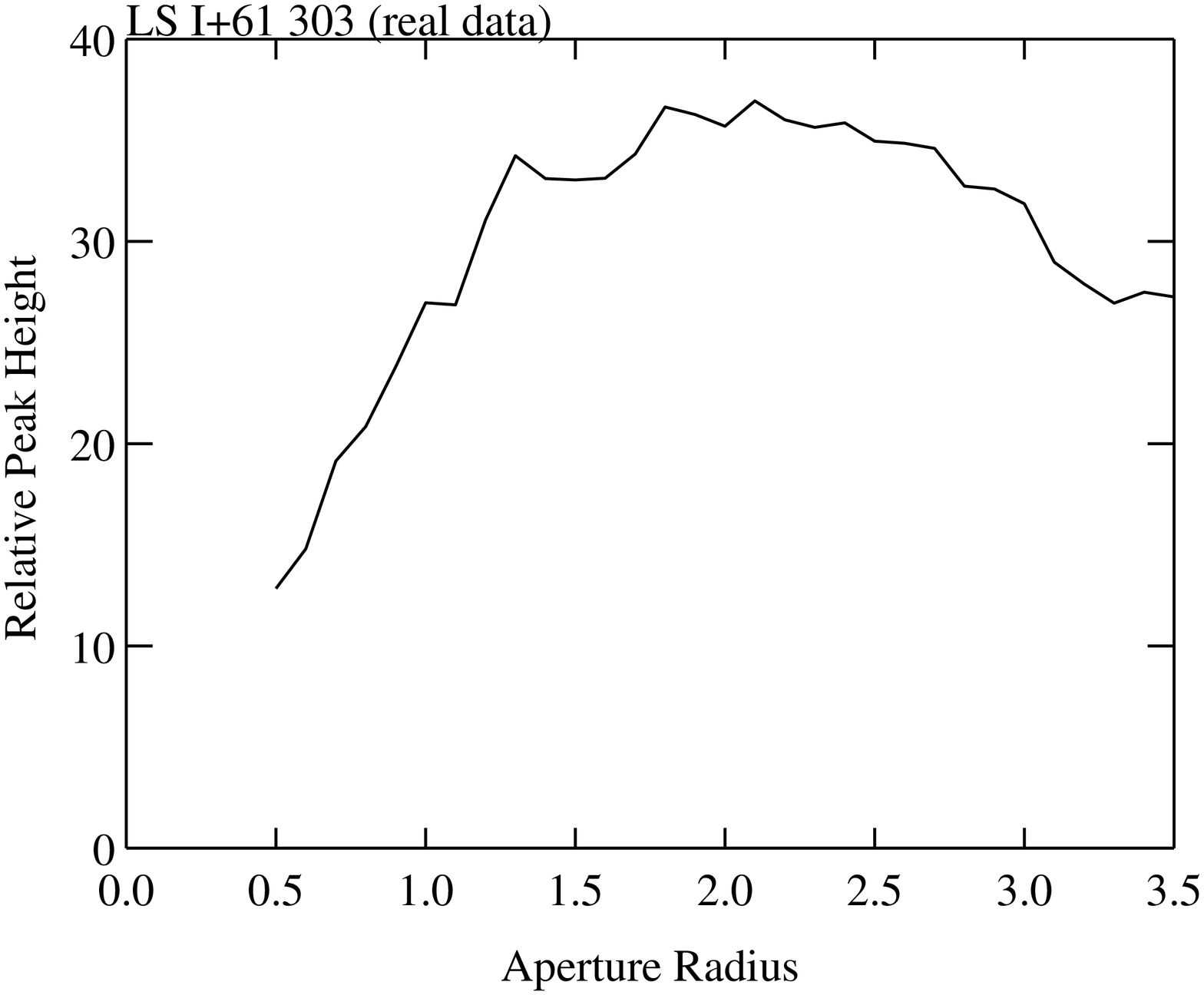}
\caption{Strength of the modulation at the orbital period of \lsi61\
in the power spectrum as a function of aperture size.}
\label{lsi61-app}
\end{figure}

\begin{figure}
\includegraphics[width=65mm,angle=270]{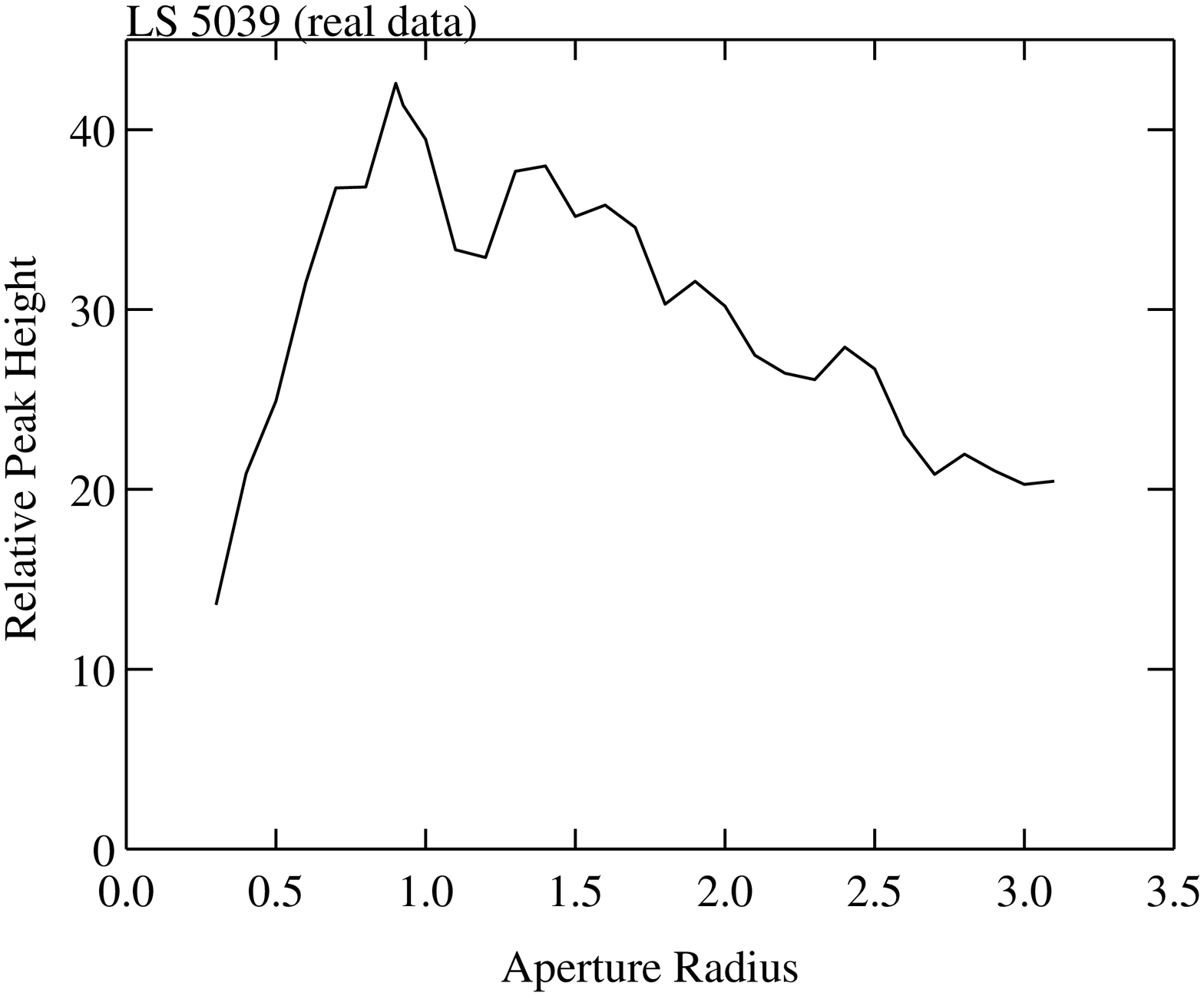}
\caption{Strength of the modulation at the orbital period of LS 5039
in the power spectrum as a function of aperture size.}
\label{ls5039-app}
\end{figure}

\begin{figure}
\includegraphics[width=65mm,angle=270]{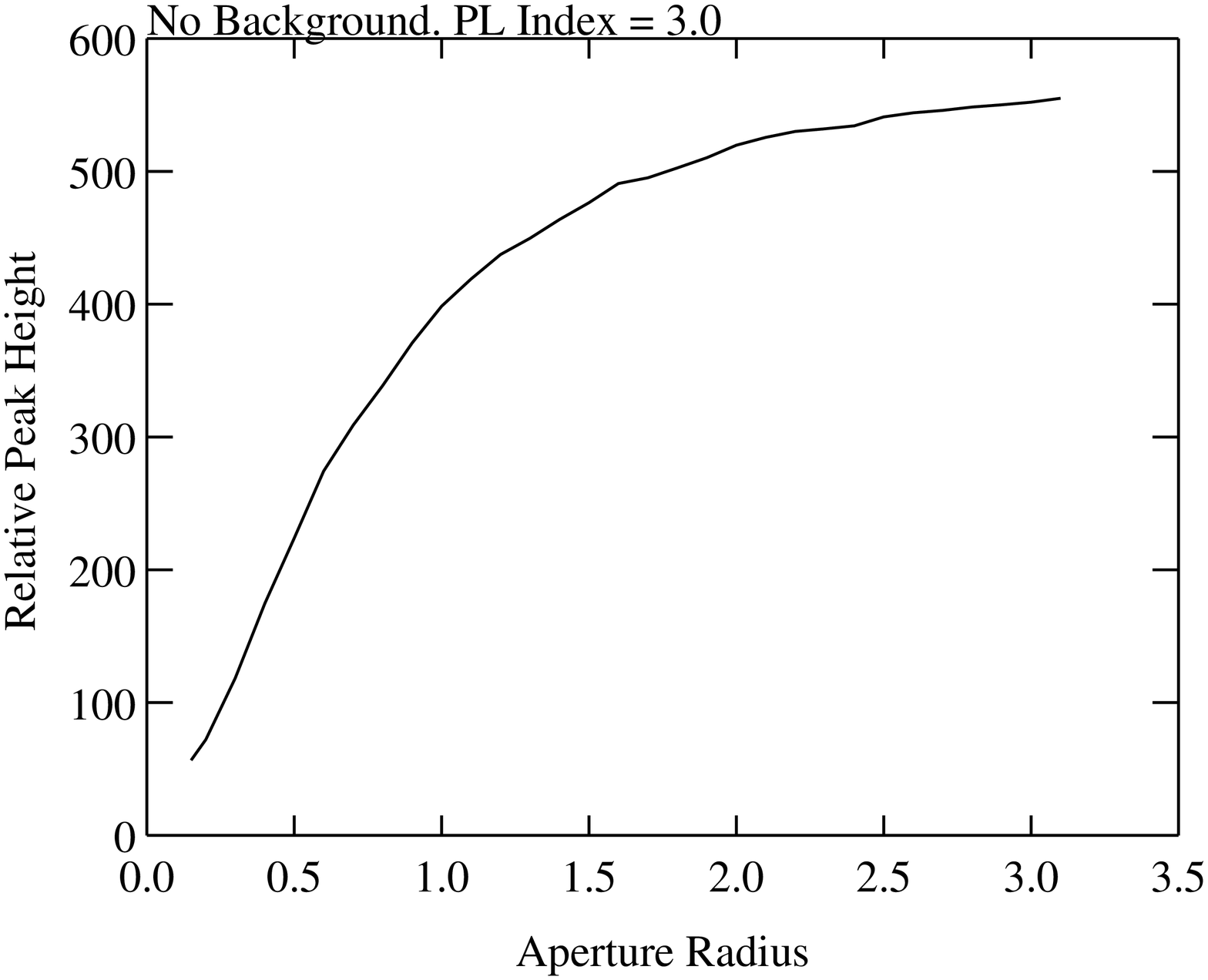}
\caption{Strength of the modulation at the orbital period 
as a function of aperture size for a simulated light curve of
a source with a power-law spectrum of index 3 and no background.}
\label{sim1-app}
\end{figure}

\begin{figure}
\includegraphics[width=65mm,angle=270]{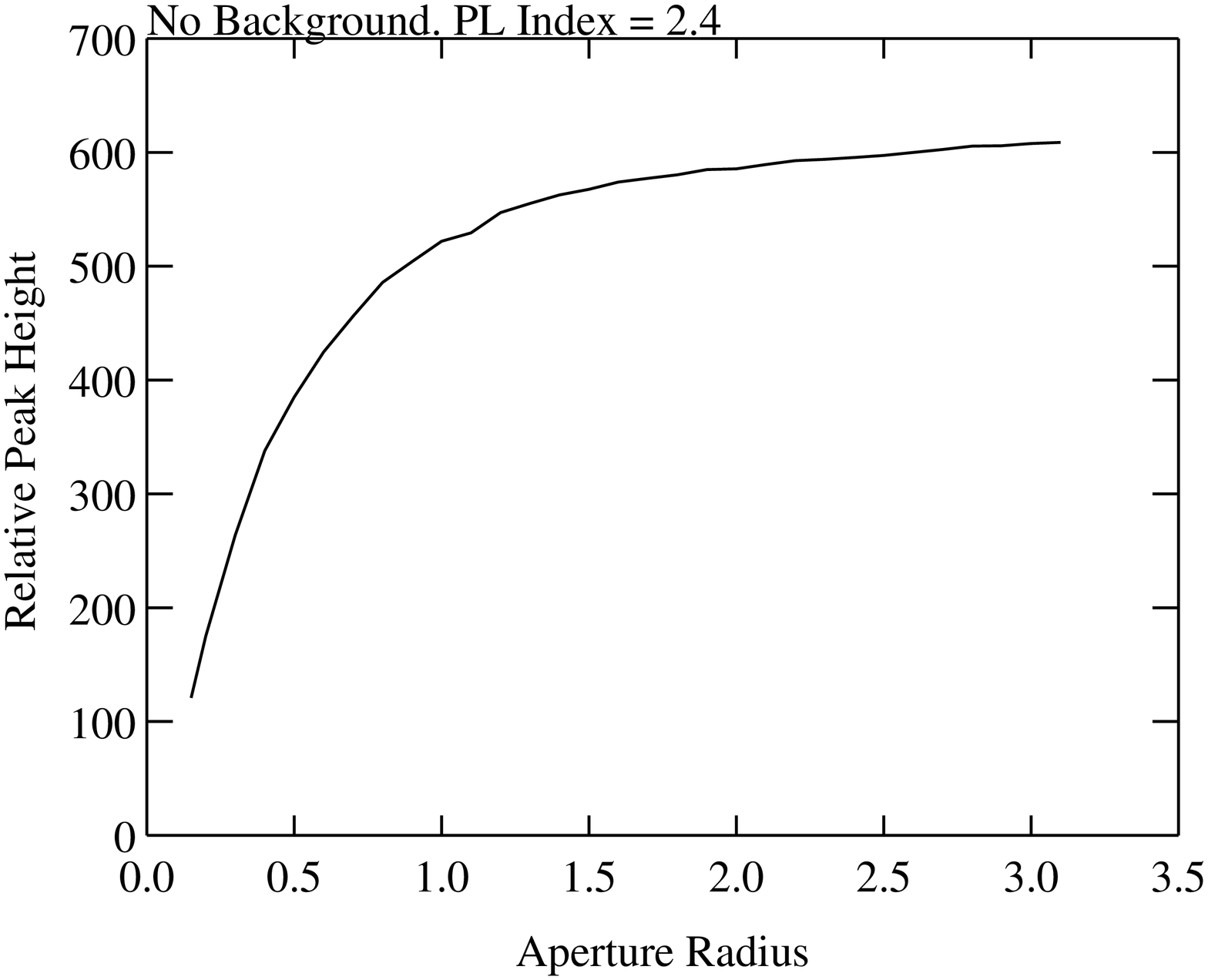}
\caption{Strength of the modulation at the orbital period 
as a function of aperture size for a simulated light curve of
a source with a power-law spectrum of index 2.4 and no background.}
\label{sim2-app}
\end{figure}

\begin{figure}
\includegraphics[width=65mm,angle=270]{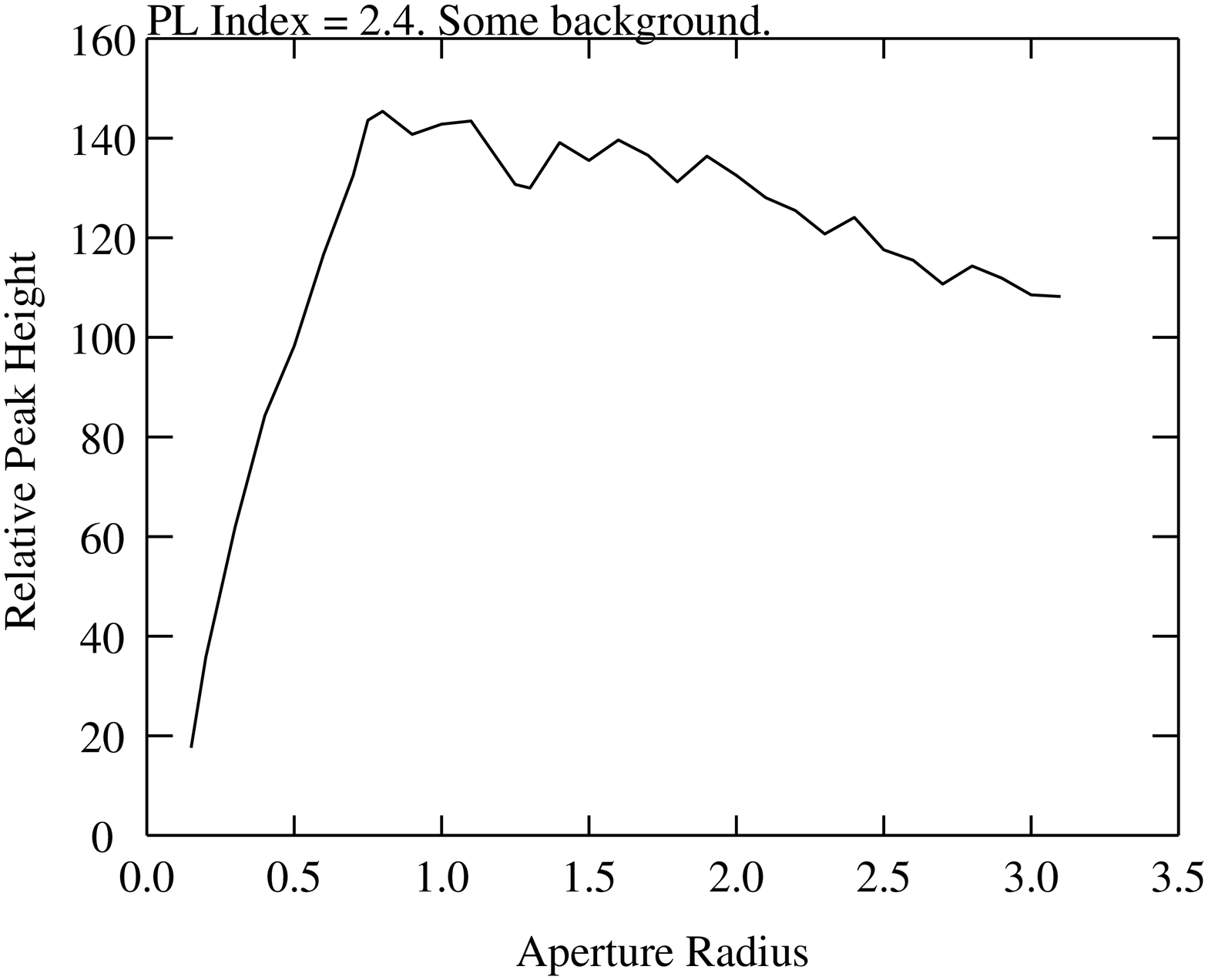}
\caption{Strength of the modulation at the orbital period 
as a function of aperture size for a simulated light curve of
a source with a power-law spectrum of index 2.4 and isotropic
background.}
\label{sim3-app}
\end{figure}

\begin{figure}
\includegraphics[width=65mm,angle=270]{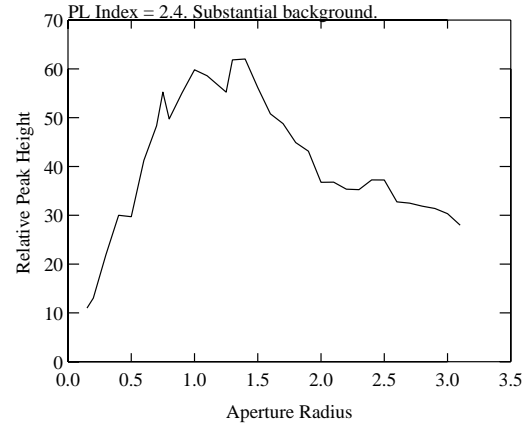}
\caption{Strength of the modulation at the orbital period 
as a function of aperture size for a simulated light curve of
a source with a power-law spectrum of index 2.4 and isotropic background
at a higher level than in Figure \ref{sim3-app}.}
\label{sim4-app}
\end{figure}

\section{Examples of Aperture Dependency}
Figures \ref{lsi61-app} to \ref{sim4-app} show the dependence of a peak in the
power spectrum at the orbital period for both observations of
modulation in actual sources (\lsi61 and LS 5039; Figures \ref{lsi61-app}
and \ref{ls5039-app}) and for
simulations using \texttt {gtobssim} (Figures \ref{sim1-app} to \ref{sim4-app}). 
These illustrate some of the dependencies
of optimal aperture size on source spectrum and the level of
background.

\section{Power Spectrum Weighting}
For Fermi light curves the exposure of time bins is not necessarily
uniform. For example, if the time bin size is a lot shorter than the
survey repeat time, then there can be extreme differences in the
exposure of the time bins. Scargle (1989) noted that the effect of
unequally weighted data points in the calculation of a periodogram
could be understood by considering the combination of points that
coincide in time. This leads to an analogy between a weighted power
spectrum and the weighted mean. However, although the procedure to
calculate a weighted periodogram is straightforward, it is important
that the correct weights are chosen. For example, if source
variability is significant compared to the typical errors on data
points then a procedure based on the semi-weighted mean of 
Cochran (1937, 1954)
is more
generally applicable (Corbet et al. 2007a, b).

For LAT light curves the number of photons in a time bin is typically
extremely small. If a weight is chosen based on the number of counts
in a bin, then bins with the same exposure could receive different
weighting.  The number of photons for the same underlying count rate
will exhibit variations due to Poisson fluctuations and a weighting
based on this would also have noise due to this shot noise. For LAT
light curves we therefore use weighting factors based on the relative
exposure of the time bins. We calculate the mean count rate of all
bins and then calculate the number of counts expected in each bin
based on its exposure. We adopt an effective error for each bin which
is the square root of the number of predicted photons divided by the
exposure.

\begin{figure}
\includegraphics[width=150mm]{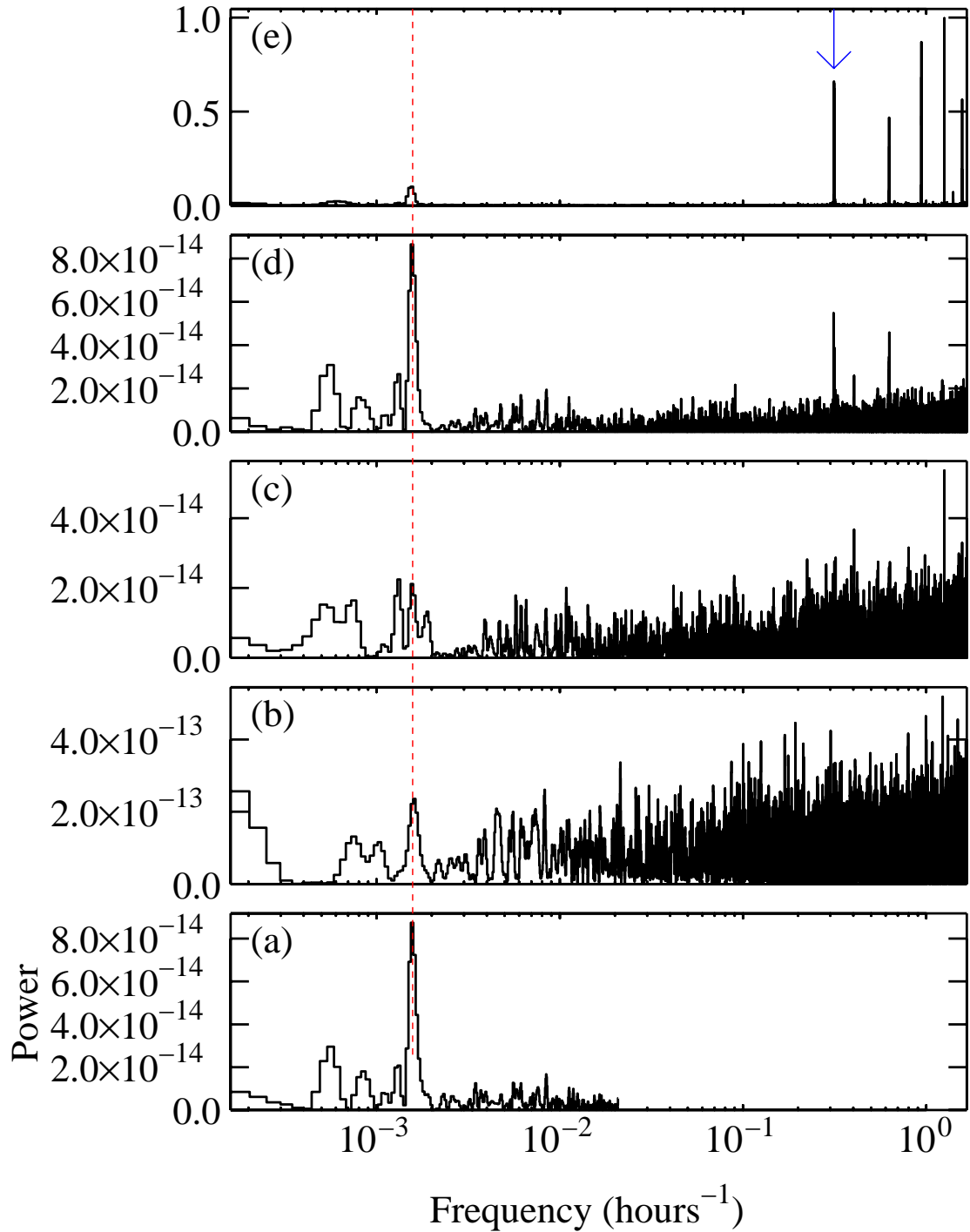}
\caption{
Power spectra of \lsi61\ obtained in five different ways. (a):
power spectrum of a 1 day time resolution light curve without
weighting. (b): power spectrum of a 1ks time resolution light curve
without weighting. (c): power spectrum of a 1ks light curve with 
``error weighting''. {\bf (d): power spectrum of 1ks light curve with 
``exposure weighting''.} (e): power spectrum of the raw number of
counts in each time bin, not corrected or weighted by exposure. The
dashed red line shows the frequency corresponding to the orbital
period of \lsi61\ and the blue arrow in the top panel indicates
Fermi's sky survey repeat frequency.}
\label{power-plot}
\end{figure}

The gain in sensitivity using weighting is illustrated in Figure
\ref{power-plot} where we show the power spectrum of the light curve of \lsi61\
(Abdo et al. 2009a) calculated in several different ways. The
bottom panel shows the power spectrum of a one day time resolution
light curve without using weighting. In this case the exposure of the
time bins is fairly uniform and the orbital period of the system is
readily detected. In the other panels are shown power spectra of \lsi61\
using 1,000 s time bins. It can be seen that without weighting
the periodicity is not detectable. Even if weighting based on the
number of photons in a time bin (``error weighting'') is used
negligible gain in signal at the orbital period is obtained. {\bf For
weighting based on the exposure (``exposure weighting'') the signal
strength is comparable to that obtained with one day binning, but
provides higher time resolution.} We note that although the exposure
weighted power spectrum of the count rate has some mathematical
similarity with just taking the power spectrum of the number of counts
in a time bin, that procedure produces large artifacts at Fermi's
rocking period and its harmonics. The use of exposure weighting
has been employed to enable the detection of the 4.8 hour orbital period of
Cygnus X-3 (Abdo et al. 2009b, Corbel 2009).

\clearpage
\bigskip % extra skip inserted
% Create the reference section using BibTeX:
%\bibliography{basename of .bib file}

\begin{thebibliography}{99} % Use for 10-99 references
\bibitem{Abdoa}
Abdo, A., et al. 2009a, ApJ, 701, L123
\bibitem{Abdob}
Abdo, A., et al. 2009b, Science, 326, 1512
\bibitem{Cochran37}
Cochran, W.G. 1937, Supplement to the Journal of the Royal Statistical Society,
4, 102
\bibitem{Cochran54}
Cochran, W.G. 1954, Biometrics, 10, 101
\bibitem{Corbel}
Corbel, S. 2009, Proceedings of the 2009 Fermi Symposium,
eConf Proceedings C091122
\bibitem{Corbeta}
Corbet, R.H.D., Markwardt, C.B., \& Tueller, J. 2007a, ApJ, 655, 458
\bibitem{Corbetb}
Corbet, R.H.D., Markwardt, C., Barbier, L., Barthelmy, S., 
Cummings, J., Gehrels, N., Krimm, H., 
Palmer, D., Sakamoto, T., Sato, G., \& Tueller, J.  2007b, 
Progress of Theoretical Physics Supplement, 169, 200
\bibitem{Holder}
Holder, J. 2009, Proceedings of the 2009 Fermi Symposium,
eConf Proceedings C091122
(arXiv:0921.4781) 
\bibitem{Howell}
Howell, S.B. 1989, PASP, 101, 616
\bibitem{Kerr}
Kerr, M. 2009, Proceedings of the 2009 Fermi Symposium,
eConf Proceedings C091122
\bibitem{Naylor}
Naylor, T. 1998, MNRAS, 296, 339
\bibitem{Scargle}
Scargle, J.D. 1989, ApJ, 343, 874


%\bibitem{exampl-ref}
%A.N. Other, ``A Very Interesting Paper'', EPAC'96, Sitges, June
%1996.

%\bibitem{templates-ref}
%http://www.cern.ch/accelconf/templates.html

\end{thebibliography}
%\begin{thebibliography}{9}   % Use for  1-9  references

\end{document}